\documentclass{sf2a-conf2018}
\usepackage{graphicx}
\usepackage{hyperref}
\usepackage[]{natbib}  
\usepackage{epstopdf}

\def\BibTeX{{\rm B\kern-.05em{\sc i\kern-.025em b}\kern-.08em
    T\kern-.1667em\lower.7ex\hbox{E}\kern-.125emX}}
\bibpunct{(}{)}{;}{a}{}{,}  

\usepackage{amsmath,amssymb,bm,mathtools,xcolor} 

\renewcommand{\vec}{\bm}
\newcommand{\cross}{\!\boldsymbol{\times}\!}
\newcommand{\bcdot}{\!\boldsymbol{\cdot}\!}

\newcommand{\vB}{\mathbf{B}}

\newcommand{\vbr}{\mathbf{r}}

\newcommand{\vY}{\mathbf{Y}}

\newcommand{\bxi}{\boldsymbol{\xi}}

\newcommand{\grad}{\boldsymbol{\nabla}}
\newcommand{\dvg}{\boldsymbol{\nabla}\!\bcdot}
\newcommand{\curl}{\boldsymbol{\nabla}\!\!\boldsymbol{\times}\!\!}
\newcommand{\cnabla}{\!\boldsymbol{\cdot}\!\boldsymbol{\nabla}}

\begin{document}

\TitreGlobal{SF2A 2018}


\title{Perturbations of Stellar Oscillations:\\ a General Magnetic Field}
\runningtitle{Stellar Oscillations \& Magnetic Fields}
\author{K.~C. Augustson}\address{AIM, CEA, CNRS, Universit\'{e} Paris-Saclay, Universit\'{e} Paris Diderot, Sarbonne Paris Cit\'{e}, F-91191 Gif-sur-Yvette Cedex, France}
\author{S. Mathis$^1$}

\setcounter{page}{237}

\maketitle


\begin{abstract}
  The purpose of this short paper is to give a summary of the derivation of a perturbative method to assess the impact of an
  arbitrary 3D magnetic field configuration, whose energy is everywhere small relative to the gravitational binding energy,
  on the form and value of the eigenfrequencies and eigenfunctions of stellar oscillations. 
\end{abstract}



\section{Magnetic Perturbations of Stellar Oscillations}

The starting point of the perturbative method is the unperturbed spheroidal modes of nonrotating and unmagnetized media
\citep[e.g.,][]{unno89}. These are the well-known f, g, and p-modes, whose base properties follow from the
spherically-symmetric thermodynamic state and gravitational potential present in nonrotating and unmagnetized
stars. Being spheroidal, these modes lack a toroidal wave component that is found when including the effects of a
magnetic field.

In the following, the principles behind linear waves in a global geometry that captures the stratification of a stellar
or planetary body and its magnetic field are established in a compact form utilizing tensorial spherical harmonics. As a
first step, the dynamics are considered in the inviscid limit, ignore rotation, thermal conduction, and omit any flows.
Electromagnetic effects are included in the magnetohydrodynamic limit, neglecting magnetic diffusion.  Moreover, the
effects of the magnetic field are assumed to be perturbative, meaning that the zeroth-order background state consisting
of the gravitational potential, pressure, and density are spherically symmetric.  To date, the perturbative impact of
specific magnetic field configurations on the eigenfrequencies have been assessed.  For instance, \citet{gough90}
considered axisymmetric magnetic fields, \citet{shibahashi93} considered an oblique dipole, and \citet{kiefer18}
considered axisymmetric toroidal magnetic fields. Here, however, no geometrical constraints are placed on the magnetic
field, namely it may be nonaxisymmetric as observed at the surface of many stars \citep[e.g.,][]{donati09,brun17}. The
modes may be decomposed onto a Fourier basis in time with $\bxi = \sum_k\bxi_k e^{i\omega_k t}$.  Thus, the linearized
and nondimensionalized MHD system of equations forms an eigenvalue problem for the eigenfrequency $\omega_k$ and
eigenfunction $\bxi_k$ that span its Hilbert space under the inner product over the spatial volume
$\int_Vd^3\vbr \rho\bxi_k^*\cdot\bxi_j$, where the star denotes complex conjugation and $\rho$ is the density, with

\vspace{-0.25truein}
\begin{center}
  \begin{align}
    &\omega_k^2\rho\bxi_k + \vec{H}\left(\bxi_k\right) + \delta\vec{L}\left(\bxi_k\right) = 0, \label{eqn:eigenvalues}
\end{align}
\end{center}

\noindent where $\vec{H}$ encapsulates the linear terms arising from the hydrostatic state and $\vec{L}$ captures the
Lorentz force as

\vspace{-0.25truein}
\begin{center}
  \begin{align}
    \vec{H}\left(\bxi\right) &= -\frac{\dvg{\left(\rho\bxi\right)}}{\rho}\grad p\!+\!\grad\!\left[\bxi\cnabla p \!+\! \rho c_s^2 \dvg{\bxi}\right]
    +\!\rho\grad\int_Vd^3\vbr'\frac{\grad_{\vbr'} \bcdot \left[\rho(\vbr')\bxi(\vbr')\right]}{\left|\vbr-\vbr'\right|}\label{eqn:eigenH} \\ 
    \vec{L}\left(\bxi\right) &= \left(\curl\vB\right)\!\cross\!\left(\curl{\left(\bxi\cross\vB\right)}\right)
    \!+\!\left(\curl \curl{\left(\bxi\cross\vB\right)}\right)\!\cross\!\vB +\frac{\dvg{\left(\rho\bxi\right)}}{\rho}\left(\curl\vB\right)\cross\vB.\label{eqn:eigenL}
\end{align}
\end{center}

\noindent where $\vB$ is the magnetic field, $c_s$ is the sound speed, $\vbr$ is the position vector, and $p$ is the
pressure. Note however that the eigenfrequencies and eigenvectors of the zeroth-order problem are degenerate in $m$ due
to its spherical symmetry. There may also be accidental degeneracies and moreover the eigenvalue problem is nonlinear;
these issues are addressed in depth in \citet{augustson18}.

Instead, a set of approximate eigenfunctions and their associated eigenfrequencies may be constructed by expanding the
true eigenfrequencies and eigenfunctions in a formal series with respect to the ratio of the nondimensionalizing
Alfv\'{e}n frequency and the Lamb frequency. This permits the perturbative classification and nondimensionalization of
the various terms in Equation \ref{eqn:eigenvalues} with the independent perturbative control parameter being
$\delta = G^{-1} M^{-2} B^2R^4/(4 \pi)$, where $B$ is a fiducial value of the magnetic field, $G$ is the gravitational
constant, $M$ is the mass contained in the domain, and $R$ is the radial extent of the domain.  Thus, the eigenfunctions
with zero eigenfrequency of Equation \ref{eqn:eigenvalues} may be linearized with respect to $\delta$, capturing the
effect of the Lorentz force and the modification of the background stratification that it induces.  This yields two sets
of equations: the perturbative equations for the zeroth-order spherically symmetric background and for the first order
magnetic perturbations to it.  The largest effect of the magnetic field will be to modify the equipotential surfaces and
also the radius of the photosphere, altering the eigenfunctions and eigenfrequencies. To account for the asphericity of
the volume and the photospheric boundary, the coordinate system can be adjusted to accomodate the magnetically-induced
spherical symmetry breaking.  The coordinate map from spherical coordinates $(r,\theta,\varphi)$ to an alternate coordinate
system $(x,\theta,\varphi)$ yields a new effective radius $x(r,\theta,\varphi)$ which may be expressed to first-order in
$\delta$ as $r/x = 1\!+\! \delta \sum_{\ell,m}\!\!h_{B,\ell}^m\left(x\right)Y_\ell^m\left(\theta,\varphi\right)$, where
$h_B$ encapsulates the local changes in the map due to the magnetic field. The solutions for $h_B$ depend in turn upon
the modified density and pressure \citep{gough90}. This ensures that a fixed radius $x$ is equivalent to an
equipotential surface. Yet, since the horizontal coordinates remain unchanged, spherical harmonics still form an
orthonormal basis on the space.

In the perturbation analysis, the eigenfrequencies and eigenfunctions are expanded in a series with respect to
$\delta$. Thus, at each order of $\delta$, they need to be defined. For the zeroth-order hydrostatic eigenvalue problem,
one has that $\omega_0^2\rho_0\bxi_0 + \vec{H}_0\left(\bxi_0\right)=0$. Since the eigenfunctions are vectors, the spin
vector harmonics (SVH) provide a complete orthonormal basis \citep[e.g.,][]{varshalovich88}.  Upon substituting the full
series, it is easily seen that the previous equation generates a set of differential equations for radially-dependent
coefficients at each of the quantizing integers $n$, $\ell$, and $m$ which denote respectively the radial order and
spherical harmonic degree and the azimuthal order.  The resulting eigenfunctions are
$\bxi_{n,\ell,0}^m = \sum_{j=\ell-1}^{\ell+1} \xi_{n,\ell,j,0}^m(x)\vY_{\ell,j}^m$, where the $\vY$ are the SVH, $\ell$
is their spherical harmonic degree, $m$ is their azimuthal order, and $j$ is their total angular momentum quantum number
$j=\ell+s$, where $s=1$ for the SVH.  The magnetic field is similarly expanded on this basis.  The equation of motion
may also be expanded with respect to $\delta$, where the time-independent terms have been eliminated using the zero
eigenvalue equations.  Due to the linearity of the problem, the eigenvalue problem for the first-order perturbation may
be cast into the following form

\vspace{-0.25truein}
\begin{center}
  \begin{align}
    &\rho_B \omega_0^2 \bxi_0 + 2\rho_0 \omega_0 \omega_B \bxi_0 + \rho_0 \omega_0^2\bxi_B +\vec{H}_0\left(\bxi_B\right) +\vec{H}_B\left(\bxi_0\right)+ \vec{L}_0\left(\bxi_0\right) = 0, \label{eqn:eigenfirst}
\end{align}
\end{center}

\noindent where the zeroth-order equation has been subtracted and $H_B$ encapsulates the first-order changes in the H
operator due to the coordinate transformation that accounts for the magnetically-induced asphericity. Hence, to
illustrate the simplicity of the SVH in building the frequency splittings, one may approximately solve Equation
\ref{eqn:eigenfirst} by taking the dot product with $\bxi_{0,n,l,m}$ and integrating over the space, yielding an
expression for the perturbation of eigenfrequencies $\omega_{B}$:

\begin{equation}
\omega_{B;n,l}^m=-\frac{1}{2\omega_{0;n,l}\mathcal{I}_{n,l}^m}\int_{V}d^3\vbr\bxi_{0,n,l}^{*m}\bcdot\left(\omega_{0,n,l}^2\rho_B\bxi_{0,n,l}^m+H_B\left(\bxi_{0,n,l}^m\right)+L_0\left(\bxi_{0,n,l}^m\right)\right)=-\frac{\mathcal{B}_{n,l}^m}{2\omega_{0;n,l}\mathcal{I}_{n,l}^m},
\end{equation}

\noindent where
$\mathcal{I}_{n,l}^m=\int_{V}d^3\vbr\rho_0|\bxi_{0,n,l}^m|^2=\sum_j\int_0^Rdx x^2\rho_0\left(x\right)|\xi_{0,n,l,j}^m\left(x\right)|^2$
is the mode inertia.  If one focuses on the terms with the largest magnitude, the Lorentz force, then the magnetically-induced
splittings may be written as

\begin{align}
&\mathcal{B}_{n,l}^m\!=\!\!\!\!\!\!\!\sum_{\substack{j,j',j'' \\ l_k, m_k, j_k}}
\!\!\int_{0}^{R}\!\!\!\mathrm{d}x x^2\mathcal{J}_{\substack{l_1,m_1,j_1 \\ l_2, m_2, j_2}}^{l, m, j} \xi_{n,l,j}^{*\, m}
\!\!\left[\mathcal{J}_{\substack{l_3, m_3, j_3 \\ l_4, m_4, j_4}}^{l_2, m_2, j'} B_{l_1,j_1}^{m_1} \mathcal{C}_{l_2, j_2}^{2, j'}\!\!\left(\xi_{n,l_3,j_3}^{m_3}B_{l_4, j_4}^{m_4}\right) +
\mathcal{J}_{\substack{l_3, m_3, j_3 \\ l_4, m_4, j_4}}^{l_1, m_1, j'}\!\!\left(\mathcal{C}_{l_1, j_1}^{1, j'}\xi_{n,l_3,j_3}^{m_3}B_{l_4,j_4}^{m_4}\right)\!\!\left(\mathcal{C}_{l_2, j_2}^{1, j''} B_{l_2,j''}^{m_2}\right)\!\right]\!\!\nonumber\\ 
&\qquad +\!\!\!\!\!\sum_{\substack{j,j' \\ l_k, m_k, j_k}}\!\!\!\!\left(-1\right)^{l+j+m_1}\mathcal{K}_{\substack{l,m,j \\ l_1, m_1, j_1}}^{l_4, m+m_1}\mathcal{J}_{\substack{l_2,m_2,j_2 \\ l_3, m_3, j_3}}^{l_1, m_2+m_3, j_1}
\!\!\int_{0}^{R}\!\!\frac{\mathrm{d}x x^2}{\rho_0\left(x\right)} \!\!\left[\xi_{n,l,j}^{*\, m}
\left(\mathcal{C}_{l_2, j_2}^{1, j'}B_{l_2,j'}^{m}\right)\!\! B_{l_3,j_3}^{m_3}\!\!\left(\mathcal{D}_{l_4,j_4}^{m+m_1}\rho_0\left(x\right)\xi_{n,l_4,j_4}^{m+m_1}\right)\!\right]\!\!.\label{M11}
\end{align}

\section{Matrix Operator Formalism}

\noindent In this effective appendix, we provide the key mathematical objects that permit the computation of the above
frequency splitting (Equation \ref{M11}). The vector cross product coefficient
$\mathcal{J}_{\substack{l_{1},m_{1},j_{1} \\ l_{2},m_{2},j_{2}}}^{l,m_{1}+m_{2},j}$ is given by

\begin{align}
&\mathcal{J}_{\substack{l_{1},m_{1},j_{1} \\ l_{2},m_{2},j_{2}}}^{l,m_{1}+m_{2},j}=(-1)^{\left(j_{1}-j_{2}+m_{1}+m_{2}\right)}\sqrt{\frac{3\left(2j+1\right)\left(2l+1\right)}{2\pi}} \mathcal{N}\!\!
\left\{\!\!
\begin{array}{ccc}
l_{1} & l_{2} & l \\ 
j_{1} & j_{2} & j \\ 
1 & 1 & 1
\end{array}\!\!
\right\}\!\! \left(\!\!
\begin{array}{ccc}
l_{1} & l_{2} & l \\ 
m_{1} & m_{2} & -\left(m_{1}+m_{2}\right)
\end{array}\!\!
\right)
\!\! \left(\!\!
\begin{array}{ccc}
j_{1} & j_{2} & j \\ 
0 & 0 & 0
\end{array}\!\!
\right)\!\!. 
\label{J}
\end{align}

\noindent Likewise the vector dot product coefficient $\mathcal{K}_{\substack{l_1,m_1,j_1 \\ l_2, m_2, j_2}}^{l, m_1+m_2}$ for the SVH is given by

\begin{align}
&\mathcal{K}_{\substack{l_1,m_1,j_1 \\ l_2, m_2, j_2}}^{l, m_1+m_2}=(-1)^{\left(j_{1}+l_{1}+l+m_1+m_2\right)}\frac{\mathcal{N}}{\sqrt{4\pi\left(2l+1\right)}} \!\!
\left[\!\!
\begin{array}{ccc}
l_{1} & l_{2} & l \\ 
j_{1} & j_{2} & 1
\end{array}\!\!
\right]\!\! \left(\!\!
\begin{array}{ccc}
j_{1} & j_{2} & l \\ 
m_{1} & m_{2} & -\left(m_{1}+m_{2}\right)
\end{array}\!\!
\right)
\!\! \left(\!\!
\begin{array}{ccc}
l_{1} & l_{2} & l \\ 
0 & 0 & 0
\end{array}\!\!
\right)\!\!, 
\label{K}
\end{align}

\noindent with the normalization
$\mathcal{N}=\sqrt{(2l_{1}+1)(2j_{1}+1)}\sqrt{(2l_{2}+1)(2j_{2}+1)}$. The $\left(...\right)$, the
$\left[...\right]$, and the $\left\{...\right\}$ denote respectively the 3j, the 6j, and the 9j symbols of the Racah-Wigner
algebra \citep{varshalovich88}. The curl and double curl can be recast into matrix operators acting on the SVH basis as

\vspace{-0.25truein}
\begin{center}
  \begin{align}
    &\curl{\vec X} = i\sum_{\substack{l_x, m_x \\ j_x, j_y}}
    \!\mathcal{C}_{l_x,j_x}^{1,j_y}X_{l_x,j_y}^{m_x}\mathbf{Y}_{l_x,j_x}^{m_x},
  \end{align}
\end{center}

\noindent with the operator being

\vspace{-0.25truein}
\begin{center}
  \begin{align}
    \mathcal{C}_{l_x,j_x}^{1,j_y} = A_{j_x,j_y}^{l_x} \frac{\partial}{\partial r} +\frac{B_{j_x,j_y}^{l_x}}{r},
  \end{align}
\end{center}

\noindent whose coefficient rotation matrices are given by

\vspace{-0.25truein}
\begin{center}
  \begin{align}
    A_{j_x, j_y}^{l} &=\left[\arraycolsep=1.4pt\def\arraystretch{1}\begin{array}{ccc} 
        0 & \sqrt{\frac{l}{2l+1}} & 0 \\
       \sqrt{\frac{l}{2l+1}} & 0 & \sqrt{\frac{l+1}{2l+1}} \\
        0 & \sqrt{\frac{l+1}{2l+1}} & 0 \\
      \end{array}\right], \qquad\qquad
    B_{j_x, j_y}^{l} =\left[\arraycolsep=1.4pt\def\arraystretch{1}\begin{array}{ccc} 
        0 & -\sqrt{\frac{l^3}{2l+1}} & 0 \\
        \sqrt{\frac{l\left(l+2\right)^2}{2l+1}} & 0 & -\sqrt{\frac{\left(l+1\right)\left(l-1\right)^2}{2l+1}} \\
        0 & \sqrt{\frac{\left(l+1\right)^3}{2l+1}} & 0 \\
      \end{array}\right]\!\!.
  \end{align}
\end{center}

\noindent Likewise, one can expand the double curl by applying the above equation twice, with

\vspace{-0.25truein}
\begin{center}
  \begin{align}
    &\curl{\curl{\vec X}} = -\sum_{\substack{l_x, m_x \\ j_x, j_y}}\!\mathcal{C}_{l_x,j_x}^{2,j_y}X_{l_x,j_y}^{m_x}\mathbf{Y}_{l_x,j_x}^{m_x},
  \end{align}
\end{center}

\noindent with the operator being

\vspace{-0.25truein}
\begin{center}
  \begin{align}
    \mathcal{C}_{l_x,j_x}^{2,j_y} = C_{j_x,j_y}^{l_x} \frac{\partial^2}{\partial r^2}
    +\frac{D_{j_x,j_y}^{l_x}}{r}\frac{\partial}{\partial r}+\frac{E_{j_x,j_y}^{l_x}}{r^2},
  \end{align}
\end{center}

\noindent whose coefficient rotation matrices are given by

\vspace{-0.25truein}
\begin{center}
  \begin{align}
    C_{j_x, j_y}^{l} &=\left[\arraycolsep=1.4pt\def\arraystretch{1}\begin{array}{ccc} 
        \frac{l}{2l+1} & 0 & \frac{\sqrt{l\left(l+1\right)}}{2l+1} \\
       0 & 1 & 0 \\
       \frac{\sqrt{l\left(l+1\right)}}{2l+1} & 0 & \frac{l+1}{2l+1} \\
      \end{array}\right], \quad
    D_{j_x, j_y}^{l} =
      \left[\arraycolsep=1.4pt\def\arraystretch{1}\begin{array}{ccc} 
        \frac{2l}{2l+1} & 0 & -\frac{\left(2l\!-\!1\right)\sqrt{l\left(l\!+\!1\right)}}{2l+1} \\
        0 & 2 & 0 \\
        \frac{\left(2l\!+\!3\right)\sqrt{l\left(l\!+\!1\right)}}{2l+1} & 0 & \frac{2l+2}{2l+1} \\
      \end{array}\right]\!\!,\nonumber\\
    E_{j_x, j_y}^{l} &=-\frac{l\left(l+1\right)}{2l+1}
    \left[\arraycolsep=1.4pt\def\arraystretch{1}\begin{array}{ccc} 
       l+2 & 0 & -\sqrt{\frac{l\!+\!1}{l}}\left(l\!-\!1\right) \\
       0 & 2l\!+\!1 & 0 \\
       -\sqrt{\frac{l}{l\!+\!1}}\left(l\!+\!2\right) & 0 & l\!-\!1 \\
      \end{array}\right]\!\!.
  \end{align}
\end{center}

\begin{acknowledgements}{The authors acknowledge support from the ERC SPIRE 647383 grant and PLATO CNES grant at
    CEA/DAp-AIM.}\end{acknowledgements}

\bibliographystyle{aa}
\bibliography{augustson_splittings}

\end{document}